\documentstyle[12pt]{article}
\begin{document}
\setlength{\parskip}{2ex}
\setlength{\textwidth}{15cm}
\setlength{\textheight}{22.5cm}
\setlength{\oddsidemargin}{0.5cm}
\setlength{\evensidemargin}{0.5cm}
\setlength{\topmargin}{-1cm}
\makeatletter
\@addtoreset{equation}{section}
\makeatother
\renewcommand{\theequation}{\thesection.\arabic{equation}}
\newcommand {\equ}[1] {(\ref{#1})}
\def\be{\begin{equation}}
\def\ee{\end{equation}}
\def\bea{\begin{eqnarray}}
\def\eea{\end{eqnarray}}
\def\bean{\begin{eqnarray*}}
\def\eean{\end{eqnarray*}}
\def\ba{\begin{array}} \def\ea{\end{array}}
\def\6{\partial} \def\a{\alpha} \def\b{\beta}
\def\g{\gamma} \def\d{\delta} \def\ve{\varepsilon} \def\e{\epsilon}
\def\z{\zeta} \def\h{\eta} \def\th{\theta}
\def\vt{\vartheta} \def\k{\kappa} \def\l{\lambda}
\def\m{\mu} \def\n{\nu} \def\x{\xi} \def\p{\pi}
\def\r{\rho} \def\s{\sigma} \def\t{\tau}
\def\Ph{\phi} \def\ph{\varphi} \def\ps{\psi}
\def\o{\omega} \def\G{\Gamma} \def\D{\Delta}
\def\Th{\Theta} \def\L{\Lambda} \def\S{\Sigma}
\def\PH{\Phi} \def\Ps{\Psi} \def\O{\Omega}
\def\sm{\small} \def\la{\large} \def\La{\Large}
\def\LA{\LARGE} \def\hu{\huge} \def\Hu{\Huge}
\def\ti{\tilde} \def\wti{\widetilde}
\def\non{\nonumber\\}
\def\={\!\!\!&=&\!\!\!}
\def\+{\!\!\!&&\!\!\!+~}
\def\-{\!\!\!&&\!\!\!-~}
\def\id{\!\!\!&\equiv&\!\!\!}
\renewcommand{\AA}{{\cal A}}
\newcommand{\BB}{{\cal B}}
\newcommand{\CC}{{\cal C}}
\newcommand{\DD}{{\cal D}}
\newcommand{\EE}{{\cal E}}
\newcommand{\FF}{{\cal F}}
\newcommand{\GG}{{\cal G}}
\newcommand{\HH}{{\cal H}}
\newcommand{\II}{{\cal I}}
\newcommand{\JJ}{{\cal J}}
\newcommand{\KK}{{\cal K}}
\newcommand{\LL}{{\cal L}}
\newcommand{\MM}{{\cal M}}
\newcommand{\NN}{{\cal N}}
\newcommand{\OO}{{\cal O}}
\newcommand{\PP}{{\cal P}}
\newcommand{\QQ}{{\cal Q}}
\newcommand{\RR}{{\cal R}}
\newcommand{\SS}{{\cal S}}
\newcommand{\TT}{{\cal T}}
\newcommand{\UU}{{\cal U}}
\newcommand{\VV}{{\cal V}}
\newcommand{\WW}{{\cal W}}
\newcommand{\XX}{{\cal X}}
\newcommand{\YY}{{\cal Y}}
\newcommand{\ZZ}{{\cal Z}}
\newcommand{\journal}[4]{{\em #1~}#2\,(19#3)\,#4;}
\newcommand{\aihp}{\journal {Ann. Inst. Henri Poincar\'e}}
\newcommand{\hpa}{\journal {Helv. Phys. Acta}}
\newcommand{\sjpn}{\journal {Sov. J. Part. Nucl.}}
\newcommand{\ijmp}{\journal {Int. J. Mod. Phys.}}
\newcommand{\physu}{\journal {Physica (Utrecht)}}
\newcommand{\pr}{\journal {Phys. Rev.}}
\newcommand{\jetpl}{\journal {JETP Lett.}}
\newcommand{\prl}{\journal {Phys. Rev. Lett.}}
\newcommand{\jmp}{\journal {J. Math. Phys.}}
\newcommand{\rmp}{\journal {Rev. Mod. Phys.}}
\newcommand{\cmp}{\journal {Comm. Math. Phys.}}
\newcommand{\cqg}{\journal {Class. Quantum Grav.}}
\newcommand{\zp}{\journal {Z. Phys.}}
\newcommand{\np}{\journal {Nucl. Phys.}}
\newcommand{\pl}{\journal {Phys. Lett.}}
\newcommand{\mpl}{\journal {Mod. Phys. Lett.}}
\newcommand{\prep}{\journal {Phys. Reports}}
\newcommand{\ptp}{\journal {Progr. Theor. Phys.}}
\newcommand{\nc}{\journal {Nuovo Cim.}}
\newcommand{\app}{\journal {Acta Phys. Pol.}}
\newcommand{\apj}{\journal {Astrophys. Jour.}}
\newcommand{\apjl}{\journal {Astrophys. Jour. Lett.}}
\newcommand{\annp}{\journal {Ann. Phys. (N.Y.)}}
\newcommand{\Nature}{{\em Nature}}
\newcommand{\PRD}{{\em Phys. Rev. D}}
\newcommand{\MNRAS}{{\em M. N. R. A. S.}}

\title{\hu Algebraic structure of Gravity with Torsion}

\author{
\\{\la Otmar MORITSCH\thanks {Supported in part by the
"Fonds zur F\"orderung der wissenschaftlichen Forschung"
under Grant No. P9116-PHY.}~, Manfred SCHWEDA,}
\\{}
\\{\la and Silvio P. SORELLA}\thanks{Supported
in part by the "Fonds zur F\"orderung der wissenschaftlichen
Forschung", M008-Lise Meitner Fellowship.}
\\{}
\\{\sm\it Institut f\"ur Theoretische Physik}
\\{\sm\it Technische Universit\"at Wien}
\\{\sm\it Wiedner Hauptstra\ss e 8-10}
\\{\sm\it A-1040 Wien (Austria)}}
{\date{{\sm October 1993}}
\maketitle
{\centerline {\bf REF. TUW 93-24} }

\makebox[1cm]{}

\begin{abstract}
The BRS transformations for gravity with torsion are discussed by
using the Maurer-Cartan horizontality conditions. With the help of
an operator $\d$ which allows to decompose the exterior space-time
derivative as a BRS commutator we solve the Wess-Zumino consistency
condition corresponding to invariant Lagrangians and anomalies.
\end{abstract}

\setcounter{page}{0}
\thispagestyle{empty}


\newpage

\section{Introduction}

It is well-known that the search of the invariant Lagrangians and of
the anomalies corresponding to a given set of field transformations
can be done in a purely algebraic way by solving the BRS consistency
condition in the space of the integrated local field polynomials.

\noindent
This amounts to study the nontrivial solutions of the equation
\be
\label{CE}
s\D=0 \ ,
\ee
where $s$ is the nilpotent BRS operator and $\D$ is an integrated
local field polynomial.
Setting $\D=\int\AA$, condition \equ{CE} translates into the local
equation
\be
\label{LE}
s\AA+d\QQ=0 \ ,
\ee
where $\QQ$ is some local polynomial
and $d=dx^{\mu}\6_{\mu}$ denotes the
exterior space-time derivative which,
together with the BRS operator $s$, obeys to:
\be
\label{NIL}
s^{2}=d^{2}=sd+ds=0 \ .
\ee
$\AA$ is said nontrivial if
\be
\AA \not= s\hat\AA+d\hat\QQ \ ,
\ee
with $\hat\AA$ and $\hat\QQ$ local polynomials.
In this case the integral of $\AA$ on space-time, $\int\AA$,
identifies a cohomology class of the BRS operator $s$ and, according
to its ghost number, it corresponds to an invariant Lagrangian
(ghost number zero) or to an anomaly (ghost number one).

The local
equation \equ{LE}, due to the relations \equ{NIL} and to the algebraic
Poincar\'e Lemma~\cite{cotta,dragon}, is easily seen to generate a tower of
descent equations
\bea
\label{LADDER}
&&s\QQ+d\QQ^{1}=0\non
&&s\QQ^{1}+d\QQ^{2}=0\non
&&~~~~~.....\non
&&~~~~~.....\non
&&s\QQ^{k-1}+d\QQ^{k}=0\non
&&s\QQ^{k}=0 \ ,
\eea
with $\QQ^{i}$ local field polynomials.

As it is well-known since several years, these equations can be solved
by using a transgression procedure based on the so-called
{\it Russian formula}
\cite{witten,baulieu,dviolette,tmieg,bandelloni,ginsparg,tonin1,stora,brandt}.
More recently an alternative way of finding nontrivial solutions of
the ladder \equ{LADDER} has been proposed by one of the authors and
successfully applied to the study of the Yang-Mills gauge
anomalies~\cite{silvio}.
The method is based on the introduction of an operator $\d$ which
allows to express the exterior derivative $d$ as a BRS commutator,
i.e.:
\be
\label{DECOMP}
d=-[s,\d] \ .
\ee
One easily verifies that, once the decomposition \equ{DECOMP} has been
found, successive applications of the operator $\d$ on the polynomial
$\QQ^{k}$ which solves the last equation of the tower
\equ{LADDER} give an explicit nontrivial solution for the higher
cocycles $\QQ^{k-1}, ....., \QQ^{1},
\QQ,$ and $\AA$.

Let us mention that the decomposition \equ{DECOMP}
represents one of the most interesting features of the topological
field theories~\cite{schwarz,birmingham} and of the bosonic string
in the Beltrami parametrization~\cite{manfred}.
We remark also that solving the last equation of the tower
\equ{LADDER}
is a problem of local BRS cohomology instead of a modulo-$d$ one.
One sees then that, due to the operator $\d$, the study of the
cohomology of $s$ modulo $d$ is essentially
reduced to the study of the local cohomology of $s$ which, in turn,
can be systematically analyzed by using the powerfull technique of the
spectral sequences~\cite{dixon}. Actually, as
proven by~\cite{tataru}, the solutions
obtained by making use of the decomposition \equ{DECOMP} turn out to be
completely equivalent to that based on the {\it Russian formula}, i.e.
they differ only by trivial cocycles.

The aim of this paper is twofold. First, we prove that the
decomposition \equ{DECOMP} extends to the gravitational case and that
it holds also in the presence of torsion.
Second, we show that the operator $\d$ gives an elegant and
straightforward way of classifying the cohomology classes of the
gravitational BRS operator in any space-time dimension.
In particular, we shall see that eq.\equ{DECOMP} will allow for a
cohomological interpretation of the cosmological constant, of the
Einstein and the generalized torsion Lagrangians as well as of the
Chern-Simons terms and the gravitational anomalies.

The paper is a continuation of a previous work~\cite{werneck},
where the decomposition \equ{DECOMP}
was shown to hold in the case of pure Lorentz transformations involving
only the spin connection $\o$ and the Riemann tensor $R$ and without
taking into account the explicit presence of the vielbein $e$ and of
the torsion $T$.

In the following we will make use of the geometrical formalism
introduced by L. Baulieu and J. Thierry-Mieg~\cite{baulieu,tmieg} which
allows to
reinterpret the BRS transformations
as a Maurer-Cartan horizontality condition.
In particular, this formalism turns out to be very useful in the case
of gravity~\cite{baulieu,tmieg} since it naturally includes the torsion.
In addition, it allows to formulate the diffeomorphism transformations
as local translations in the tangent space by means of the introduction
of the ghost field $\h^{a}=\x^{\mu}e_{\mu}^{a}$ where $\x^{\mu}$
denotes the usual diffeomorphism ghost and $e_{\mu}^{a}$ is the
vielbein\footnote{As usual, Latin and Greek indices refer to the
tangent space and to the euclidean space-time.}.
This step, as we shall see in details, will allow to
introduce the decomposition \equ{DECOMP} in a very simple way.
Moreover, the explicit presence of the torsion T and of the
translation ghost $\h^{a}$ gives us the possibility of
introducing an algebraic BRS setup which turns out to be completely
different from that of a recent work of Brandt et al.~\cite{brandt}
where similar techniques has been used.

Let us finish this introduction by remarking that, as done
in~\cite{werneck}, we always refer to the gravitational fields,
i.e. to the
vielbein $e$, the spin connection $\o$, the Riemann tensor $R$, and
the torsion $T$ as unquantized classical fields which, when coupled
to some matter fields (scalars or fermions),
give rise to an effective action whose quantum expansion reduces to
the one-loop order.

The paper is organized as follows. In Section 2 we briefly recall
the Maurer-Cartan horizontality condition and we derive the BRS
transformations for the local Lorentz rotations and diffeomorphisms.
In Section 3 we introduce the operator $\d$ and we show how it can be
used to solve the descent equations \equ{LADDER}. Section 4 is devoted
to the study of some explicit examples and Section 5 deals with the
geometrical meaning of the decomposition \equ{DECOMP}. Some detailed
calculations are given in the final Appendices.

\section{\hspace{-0.07cm}The Maurer-Cartan horizontality condition}

The aim of this section is to derive the gravitational BRS
transformations from the Maurer-Cartan geometrical
formalism~\cite{baulieu,tmieg}.
For a better understanding of the so-called horizontality condition
let us begin by considering the simpler case of a nonabelian
Yang-Mills theory.

\subsection{The Yang-Mills case}

Denoting with $A^{a}=A^{a}_{\mu}dx^{\mu}$ and $c^{a}$ the
one form gauge connection and the zero form ghost field, for the BRS
transformations one has:
\bea
\label{BRSYM}
sA^{a}\=dc^{a}+f^{abc}c^{b}A^{c}\non
sc^{a}\=\frac{1}{2}f^{abc}c^{b}c^{c}\non
s^{2}\=0 \ ,
\eea
where $f^{abc}$ are the structure constants of the corresponding
gauge group.
As usual, the adopted graduation is given by the sum of the form degree
and of the ghost number. The fields $A^{a}$ and $c^{a}$ are both of
degree one, their ghost number being respectively zero and one.
A $p$-form with ghost number $q$ will be denoted by $\O^{q}_{p}$, its
total graduation being $(p+q)$.
The two form field strength $F^{a}$ is given by
\be
F^{a}=\frac{1}{2}F^{a}_{\mu\nu}dx^{\mu}dx^{\nu}
=dA^{a}+\frac{1}{2}f^{abc}A^{b}A^{c},
\ee
and
\be
dF^{a}=f^{abc}F^{b}A^{c},
\ee
is its Bianchi identity.
In order to reinterpret the BRS transformations \equ{BRSYM} as a
Maurer-Cartan horizontality condition we introduce the combined
gauge-ghost field
\be
\wti{A}^{a}=A^{a}+c^{a},
\ee
and the generalized nilpotent differential operator
\be
\label{GDO}
\wti{d}=d-s \ , ~~~\wti{d}^{2}=0 \ .
\ee
Notice that both $\wti{A}^{a}$ and $\wti{d}$ have degree one.
Let us introduce also the degree-two field strenght $\wti{F}^{a}$:
\be
\wti{F}^{a}=\wti{d}\wti{A}^{a}
+\frac{1}{2}f^{abc}\wti{A}^{b}\wti{A}^{c},
\ee
which, from eq.\equ{GDO}, obeys the generalized Bianchi identity
\be
\wti{d}\wti{F}^{a}=f^{abc}\wti{F}^{b}\wti{A}^{c}.
\ee
The Maurer-Cartan horizontality condition~\cite{baulieu,tmieg}
reads then
\be
\label{MCYM}
\wti{F}^{a}=F^{a}.
\ee
It is very easy now to check that the BRS transformations \equ{BRSYM}
can be obtained from the horizontality condition \equ{MCYM} by simply
expanding $\wti{F}^{a}$ in terms of the elementary fields $A^{a}$ and
$c^{a}$ and collecting the terms with the same form degree and ghost
number. In addition, let us remark also the equality
\be
\wti{d}\wti{F}^{a}-f^{abc}\wti{F}^{b}\wti{A}^{c}
=dF^{a}-f^{abc}F^{b}A^{c}.
\ee

\subsection{The gravitational case}

In order to generalize the horizontality condition \equ{MCYM}
to the gravitational case let us first specify the functional space
the BRS operator $s$ acts upon. The latter is chosen to be the space
of local polynomials which depend on the one forms
$(e^{a},\o^{a}_{~b})$, $e^{a}$ and $\o^{a}_{~b}$ being respectively
the vielbein and the spin connection
\bea
\label{OF}
e^{a} \= e^{a}_{\mu}dx^{\mu},\non
\o^{a}_{~b} \= \o^{a}_{~b\mu}dx^{\mu},
\eea
and on the two forms $(T^{a},R^{a}_{~b})$, $T^{a}$ and $R^{a}_{~b}$
denoting the torsion and the Riemann tensor
\bea
\label{TF}
T^{a} \= \frac{1}{2}T^{a}_{\mu\nu}dx^{\mu}dx^{\nu}
=de^{a}+\o^{a}_{~b}e^{b}
=De^{a},\non
R^{a}_{~b} \= \frac{1}{2}R^{a}_{~b\mu\nu}dx^{\mu}dx^{\nu}
=d\o^{a}_{~b}+
\o^{a}_{~c}\o^{c}_{~b} \ ,
\eea
where
\be
D=d+\o
\ee
is the covariant derivative.
The tangent space indices $(a,b,c,.....)$ in eqs.\equ{OF} and
\equ{TF} are referred to the group $SO(N)$, $N$ being the dimension
of the euclidean space-time.

Applying the exterior derivative $d$ to both sides of eq.\equ{TF}
one gets the Bianchi identities
\bea
\label{BI}
DT^{a} \= dT^{a}+\o^{a}_{~b}T^{b}=R^{a}_{~b}e^{b},\non
DR^{a}_{~b} \= dR^{a}_{~b}+\o^{a}_{~c}R^{c}_{~b}
-\o^{c}_{~b}R^{a}_{~c}=0 \ .
\eea

To write down the gravitational Maurer-Cartan horizontality condition
let us introduce, as done in~~\cite{baulieu,tmieg},
the local Lorentz ghost
$\th^{a}_{~b}=-\th_{b}^{~a}$
and a further ghost $\h^{a}$ with indices in the tangent space.
Both $\th^{a}_{~b}$ and $\h^{a}$ have ghost number one.

As we shall see later, the introduction of the ghost $\h^{a}$ turns
out to be quite useful since it allows to express the diffeomorphism
transformations as local translations in the flat tangent space, the
transition between the two formulations being realized by the
relation
\be
\label{ETA}
\x^{\mu} = E^{\mu}_{a} \h^{a}~~~,~~~
\h^{a} = \x^{\mu}e^{a}_{\mu} \ ,
\ee
where $\x^{\mu}$ is the usual ghost for the diffeomorphisms and
$E^{\mu}_{a}$ denotes the inverse of the vielbein
$e^{a}_{\mu}$, i.e.
\bea
e^{a}_{\mu}E^{\mu}_{b} \= \d^{a}_{b} \ ,\non
e^{a}_{\mu}E^{\nu}_{a} \= \d^{\nu}_{\mu} \ .
\eea
Proceeding now as in the previous section, one defines the nilpotent
differential operator $\wti{d}$ of degree one:
\be
\label{EXTD}
\wti{d}=d-s \ ,
\ee
and the generalized vielbein-ghost field $\ti{e}^{a}$ and the extended
spin connection
$\wti{\o}^{a}_{~b}$
\bea
\label{EVIEL}
\ti{e}^{a} \= e^{a}+\h^{a},\non
\wti{\o}^{a}_{~b} \= \widehat{\o}^{a}_{~b}+\th^{a}_{~b} \ ,
\eea
where, following~\cite{tmieg}, $\widehat{\o}^{a}_{~b}$ is given by
\be
\label{HCO}
\widehat{\o}^{a}_{~b}=\o^{a}_{~bm}\ti{e}^{m}
=\o^{a}_{~b}+\o^{a}_{~bm}\h^{m},
\ee
with the zero form $\o^{a}_{~bm}$\footnote{We remark that the zero form
$\o^{a}_{~bm}$ does not possess any symmetric or antisymmetric property
with respect to the lower indices $(bm)$.}
defined by the expansion of the one form
connection $\o^{a}_{~b\mu}$ in terms of the vielbein $e^{a}_{\mu}$,
i.e.:
\be
\label{SPINC}
\o^{a}_{~b\mu}=\o^{a}_{~bm}e^{m}_{\mu} \ .
\ee
As it is well-known, this last formula stems from the fact that the
vielbein formalism allows to transform locally the space-time indices
of an arbitrary
tensor $\NN_{\mu\nu\r\s...}$ into flat tangent space indices
$\NN_{abcd...}$ by means of the expansion
\be
\label{WORLD}
\NN_{\mu\nu\r\s...}=\NN_{abcd...}
e^{a}_{\mu}e^{b}_{\nu}e^{c}_{\r}e^{d}_{\s}...  \ .
\ee
Vice versa
\be
\label{TANG}
\NN_{abcd...}=\NN_{\mu\nu\r\s...}
E^{\mu}_{a}E^{\nu}_{b}E^{\r}_{c}E^{\s}_{d}...  \ .
\ee
According to the definition \equ{TF}, the generalized Riemann tensor
and torsion field are given by
\bea
\label{DEFTWO}
\wti{T}^{a} \= \wti{d}\ti{e}^{a}+\wti{\o}^{a}_{~b}\ti{e}^{b}
=\wti{D}\ti{e}^{a},\non
\wti{R}^{a}_{~b} \= \wti{d}\wti{\o}^{a}_{~b}+\wti{\o}^{a}_{~c}
\wti{\o}^{c}_{~b} \ ,
\eea
and are easily seen to obey the generalized Bianchi identities
\bea
\label{GBI}
\wti{D}\wti{T}^{a} \= \wti{d}\wti{T}^{a}+\wti{\o}^{a}_{~b}\ti{e}^{b}
=\wti{R}^{a}_{~b}\ti{e}^{b},\non
\wti{D}\wti{R}^{a}_{~b} \= \wti{d}\wti{R}^{a}_{~b}
+\wti{\o}^{a}_{~c}\wti{R}^{c}_{~b}
-\wti{\o}^{c}_{~b}\wti{R}^{a}_{~c}=0 \ ,
\eea
with
\be
\wti{D}=\wti{d}+\wti{\o}
\ee
the generalized covariant derivative.

We are now ready to formulate the Maurer-Cartan equations for gravity.
Following~\cite{tmieg}, these conditions state that {\it $\ti{e}$ and
all
its generalized Lorentz covariant exterior differentials can be
expanded over $\ti{e}$ with classical coefficients}, i.e.:
\bea
\label{MCG1}
\ti{e}^{a}\=\d^{a}_{b}\ti{e}^{b} \equiv horizontal,\\
\label{MCG2}
\wti{T}^{a}(\ti{e},\wti{\o})
\=\frac{1}{2}T^{a}_{mn}(e,\o)\ti{e}^{m}\ti{e}^{n}
\equiv horizontal,\\
\label{MCG3}
\wti{R}^{a}_{~b}(\wti{\o})
\=\frac{1}{2}R^{a}_{~bmn}(\o)\ti{e}^{m}\ti{e}^{n}
\equiv horizontal,
\eea
where, from eq.\equ{WORLD}, the zero forms $T^{a}_{mn}$ and
$R^{a}_{~bmn}$ are defined by the vielbein expansion of the two forms
torsion and Riemann tensor of
eq.\equ{TF},
\bea
\label{TWOFORM}
T^{a}\=\frac{1}{2}T^{a}_{mn}e^{m}e^{n},\non
R^{a}_{~b}\=\frac{1}{2}R^{a}_{~bmn}e^{m}e^{n}.
\eea
Notice also that eq.\equ{HCO} is nothing but the horizontality
condition for the spin connection expressing the fact that
$\widehat{\o}$ itself can be expanded over $\ti{e}$.

Eqs.\equ{MCG1}-\equ{MCG3} define the Maurer-Cartan horizontality
conditions for the gravitational case and, when expanded in terms of
the elementary fields
$(e^{a}, \o^{a}_{~b}, \h^{a}, \th^{a}_{~b})$, give the nilpotent BRS
transformations corresponding to the local Lorentz rotations and to
the diffeomorphism transformations.

For a better understanding of this point let us discuss in details the
horizontality condition \equ{MCG2} for the torsion.
Making use of eqs.\equ{EXTD}, \equ{EVIEL}, \equ{HCO} and of the
definition \equ{DEFTWO}, one verifies that eq.\equ{MCG2} gives
\bea
&&de^{a}-se^{a}+d\h^{a}-s\h^{a}+\o^{a}_{~b}e^{b}+\th^{a}_{~b}e^{b}\non
&&+~\o^{a}_{~b}\h^{b}+\th^{a}_{~b}\h^{b}
+\o^{a}_{~bm}\h^{m}e^{b}+\o^{a}_{~bm}\h^{m}\h^{b}\non
&&=\frac{1}{2}T^{a}_{mn}e^{m}e^{n}
+T^{a}_{mn}e^{m}\h^{n}+\frac{1}{2}T^{a}_{mn}\h^{m}\h^{n} \ ,
\eea
from which, collecting the terms with the same form degree and
ghost number, one easily obtains the BRS transformations for the
vielbein $e^{a}$ and for the ghost $\h^{a}$:
\bea
\label{BRSE}
se^{a}\=d\h^{a}+\o^{a}_{~b}\h^{b}+\th^{a}_{~b}e^{b}
+\o^{a}_{~bm}\h^{m}e^{b}-T^{a}_{mn}e^{m}\h^{n} \ ,\non
s\h^{a}\=\th^{a}_{~b}\h^{b}+\o^{a}_{~bm}\h^{m}\h^{b}
-\frac{1}{2}T^{a}_{mn}\h^{m}\h^{n} \ .
\eea
These equations, when rewritten in terms of the variable $\x^{\mu}$ of
eq.\equ{ETA}, take the more familiar form
\bea
se^{a}_{\mu}\=\th^{a}_{~b}e^{b}_{\mu}+\LL_{\x}e^{a}_{\mu} \ ,\non
s\x^{\mu}\=-\x^{\l}\6_{\l}\x^{\mu},
\eea
where $\LL_{\x}$ denotes the ordinary Lie derivative along the
direction $\x^{\mu}$, i.e.
\be
\LL_{\x}e^{a}_{\mu}=-\x^{\l}\6_{\l}e^{a}_{\mu}
-(\6_{\mu}\x^{\l})e^{a}_{\l} \ .
\ee
It is apparent now that eq.\equ{BRSE} represents the tangent space
formulation of the usual BRS transformations corresponding to local
Lorentz rotations and diffeomorphisms.

One sees then that the Maurer-Cartan horizontality conditions
\equ{MCG1}-\equ{MCG3} together with eq.\equ{DEFTWO} carry in a very
simple and compact way all the informations relative to the
gravitational gauge algebra. It is easy indeed to expand
eqs.\equ{MCG1}-\equ{MCG3} in terms of $e^{a}$ and $\h^{a}$ and work
out the BRS transformations of the remaining fields
$(\o^{a}_{~b}, R^{a}_{~b}, T^{a}, ...)$.

However, in view of the fact that we will use as fundamental variables
the zero forms $(\o^{a}_{~bm}, R^{a}_{~bmn}, T^{a}_{mn})$ rather than
the one form spin connection $\o^{a}_{~b}$ and the two forms
$R^{a}_{~b}$ and $T^{a}$, let us proceed by introducing the partial
derivative $\6_{a}$ with indices in the tangent space.
According to the formulas \equ{WORLD} and \equ{TANG}, the latter is
defined by
\be
\6_{a} \equiv E^{\mu}_{a}\6_{\mu} \ ,
\ee
and
\be
\6_{\mu} = e^{a}_{\mu}\6_{a} \ ,
\ee
so that the intrinsic exterior differential $d$ becomes
\be
d=dx^{\mu}\6_{\mu}=e^{a}\6_{a} \ .
\ee
Let us emphasize that the introduction of the operator $\6_{a}$ and
the use of the zero forms $(\o^{a}_{~bm}, R^{a}_{~bmn}, T^{a}_{mn})$
allows for a complete tangent space formulation of the gravitational
gauge algebra. This step, as we shall see later, turns out to be very
useful in the analysis of the corresponding BRS cohomology. Moreover,
as one can easily understand, the knowledge of the BRS transformations
of the zero form sector $(\o^{a}_{~bm}, R^{a}_{~bmn}, T^{a}_{mn})$
together with the expansions \equ{SPINC}, \equ{TWOFORM} and the
equation \equ{BRSE} completely characterize the transformation
law of the forms $(\o^{a}_{~b}, R^{a}_{~b}, T^{a})$.

Let us remark however that, contrary to the case of the usual
space-time derivative $\6_{\mu}$, the operator $\6_{a}$ does not
commute with the BRS operator
due to the explicit presence of the vielbein $e^{a}$
(see Appendix A for the detailed calculations).
One has:
\be
[s,\6_{m}]=(\6_{m}\h^{k}-\th^{k}_{~m}-T^{k}_{mn}\h^{n}
-\o^{k}_{~mn}\h^{n}+\o^{k}_{~nm}\h^{n})\6_{k} \ ,
\ee
and
\be
[\6_{m},\6_{n}]=-(T^{k}_{mn}+\o^{k}_{~mn}-\o^{k}_{~nm})\6_{k} \ .
\ee
Nevertheless, taking into account the vielbein transformation
\equ{BRSE}, one consistently verifies that
\be
\{s,d\}=0 \ ,~~~d^{2}=0 \ .
\ee

\subsection{BRS transformations and Bianchi identities}

Let us finish this chapter by giving, for the convenience of the
reader, the BRS transformations and the Bianchi identities which
come out from the Maurer-Cartan horizontality conditions
\equ{MCG1}-\equ{MCG3} and from eqs.\equ{DEFTWO}, \equ{GBI} for
each form sector and ghost number.

\begin{itemize}

\item {\bf Form sector two $(R^{a}_{~b}, T^{a})$}

\bea
\label{FORMTWO}
sR^{a}_{~b}\=\th^{a}_{~c}R^{c}_{~b}-\th^{c}_{~b}R^{a}_{~c}
+\o^{a}_{~ck}\h^{k}R^{c}_{~b}-\o^{c}_{~bk}\h^{k}R^{a}_{~c}\non
\+\o^{a}_{~c}R^{c}_{~bmn}e^{m}\h^{n}
-\o^{c}_{~b}R^{a}_{~cmn}e^{m}\h^{n}+(dR^{a}_{~bmn})e^{m}\h^{n}\non
\+R^{a}_{~bmn}T^{m}\h^{n}
-R^{a}_{~bkn}\o^{k}_{~m}e^{m}\h^{n}-R^{a}_{~bmn}e^{m}d\h^{n},\non
sT^{a}\=\th^{a}_{~b}T^{b}+\o^{a}_{~bk}\h^{k}T^{b}-R^{a}_{~b}\h^{b}\non
\+\o^{a}_{~b}T^{b}_{mn}e^{m}\h^{n}-R^{a}_{~bmn}e^{b}e^{m}\h^{n}
+(dT^{a}_{mn})e^{m}\h^{n}\non
\-T^{a}_{mn}e^{m}d\h^{n}+T^{a}_{mn}T^{m}\h^{n}
-T^{a}_{kn}\o^{k}_{~m}e^{m}\h^{n}.
\eea
For the Bianchi identities one has
\bea
\label{BIG}
&&dR^{a}_{~b}+\o^{a}_{~c}R^{c}_{~b}-\o^{c}_{~b}R^{a}_{~c}=0 \ ,\non
&&dT^{a}+\o^{a}_{~b}T^{b}=R^{a}_{~b}e^{b}.
\eea

\item {\bf Form sector one $(\o^{a}_{~b}, e^{a})$}

\bea
s\o^{a}_{~b}\=d\th^{a}_{~b}+\th^{a}_{~c}\o^{c}_{~b}
+\o^{a}_{~c}\th^{c}_{~b}
+(d\o^{a}_{~bm})\h^{m}+\o^{a}_{~bm}d\h^{m}\non
\+\o^{a}_{~c}\o^{c}_{~bm}\h^{m}+\o^{a}_{~cm}\h^{m}\o^{c}_{~b}
-R^{a}_{~bmn}e^{m}\h^{n},\non
se^{a}\=d\h^{a}+\o^{a}_{~b}\h^{b}+\th^{a}_{~b}e^{b}
+\o^{a}_{~bm}\h^{m}e^{b}
-T^{a}_{mn}e^{m}\h^{n}.
\eea

\item {\bf Form sector zero, ghost number zero
$(\o^{a}_{~bm}, R^{a}_{~bmn}, T^{a}_{mn})$ }

\bea
s\o^{a}_{~bm}\=-\6_{m}\th^{a}_{~b}+\th^{a}_{~c}\o^{c}_{~bm}
-\th^{c}_{~b}\o^{a}_{~cm}
-\th^{k}_{~m}\o^{a}_{~bk}-\h^{k}\6_{k}\o^{a}_{~bm} \ ,\non
sR^{a}_{~bmn}\=\th^{a}_{~c}R^{c}_{~bmn}-\th^{c}_{~b}R^{a}_{~cmn}
-\th^{k}_{~m}R^{a}_{~bkn}-\th^{k}_{~n}R^{a}_{~bmk}
-\h^{k}\6_{k}R^{a}_{~bmn} \ ,\non
sT^{a}_{mn}\=\th^{a}_{~k}T^{k}_{mn}-\th^{k}_{~m}T^{a}_{kn}
-\th^{k}_{~n}T^{a}_{mk}-\h^{k}\6_{k}T^{a}_{mn} \ .
\eea
The Bianchi identities \equ{BIG} are projected on the zero form
curvature $R^{a}_{~bmn}$ and on the zero form torsion $T^{a}_{mn}$
to give
\bea
dR^{a}_{~bmn}\=(\6_{l}R^{a}_{~bmn})e^{l}\non
\=(-\o^{a}_{~cl}R^{c}_{~bmn}-\o^{a}_{~cm}R^{c}_{~bnl}
-\o^{a}_{~cn}R^{c}_{~blm}\non
\+\o^{c}_{~bl}R^{a}_{~cmn}+\o^{c}_{~bm}R^{a}_{~cnl}
+\o^{c}_{~bn}R^{a}_{~clm}\non
\+R^{a}_{~bkn}T^{k}_{ml}+R^{a}_{~bkm}T^{k}_{ln}
+R^{a}_{~bkl}T^{k}_{nm}\non
\-R^{a}_{~bkn}\o^{k}_{lm}-R^{a}_{~bkm}\o^{k}_{nl}
-R^{a}_{~bkl}\o^{k}_{mn}\non
\+R^{a}_{~bkn}\o^{k}_{ml}+R^{a}_{~bkl}\o^{k}_{nm}
+R^{a}_{~bkm}\o^{k}_{ln}\non
\-\6_{m}R^{a}_{~bnl}-\6_{n}R^{a}_{~blm})e^{l},\non
dT^{a}_{mn}\=(\6_{l}T^{a}_{mn})e^{l}\non
\=(R^{a}_{~lmn}+R^{a}_{~mnl}+R^{a}_{~nlm}\non
\-\o^{a}_{~bl}T^{b}_{mn}-\o^{a}_{~bm}T^{b}_{nl}
-\o^{a}_{~bn}T^{b}_{lm}\non
\+T^{a}_{kn}T^{k}_{ml}+T^{a}_{km}T^{k}_{ln}
+T^{a}_{kl}T^{k}_{nm}\non
\-T^{a}_{kn}\o^{k}_{~lm}-T^{a}_{km}\o^{k}_{~nl}
-T^{a}_{kl}\o^{k}_{~mn}\non
\+T^{a}_{kn}\o^{k}_{~ml}+T^{a}_{kl}\o^{k}_{~nm}
+T^{a}_{km}\o^{k}_{~ln}\non
\-\6_{m}T^{a}_{nl}-\6_{n}T^{a}_{lm})e^{l}.
\eea
One has also the equation
\bea
d\o^{a}_{~bm}\=(\6_{n}\o^{a}_{~bm})e^{n}\non
\=(-R^{a}_{~bmn}+\o^{a}_{~cm}\o^{c}_{~bn}-\o^{a}_{~cn}\o^{c}_{~bm}\non
\+\o^{a}_{~bk}T^{k}_{mn}-\o^{a}_{~bk}\o^{k}_{~nm}
+\o^{a}_{~bk}\o^{k}_{~mn}+\6_{m}\o^{a}_{~bn})e^{n}.
\eea

\item {\bf Form sector zero, ghost number one $(\th^{a}_{~b}, \h^{a})$ }

\bea
\label{FORMZERO}
s\th^{a}_{~b}\=\th^{a}_{~c}\th^{c}_{~b}
-\h^{k}\6_{k}\th^{a}_{~b} \ ,\non
s\h^{a}\=\o^{a}_{~bm}\h^{m}\h^{b}+\th^{a}_{~b}\h^{b}
-\frac{1}{2}T^{a}_{mn}\h^{m}\h^{n}.
\eea

\item {\bf Algebra between $s$ and $d$ }

{}From the above transformations it follows:
\be
s^{2}=0 \ ,~~~d^{2}=0 \ ,
\ee
and
\be
\{s,d\}=0 \ .
\ee

\end{itemize}

\section{Decompositon of the exterior derivative}

In this section we introduce the decomposition \equ{DECOMP} and
we show how it can be used to solve the ladder \equ{LADDER}.
To this purpose let us introduce the operator $\d$ defined as
\bea
\label{DECETA}
\d\h^{a}\=-e^{a},\non
\d\ph\=0~~~{\rm for}~~~\ph=(\o, e, R, T, \th) \ .
\eea
It is easy to verify that $\d$ is of degree zero and that, together
with the BRS operator $s$, it obeys the following algebraic relations:
\be
\label{DEC}
[s,\d]=-d \ ,
\ee
and
\be
\label{EXCOM}
[d,\d]=0 \ .
\ee
One sees from eq.\equ{DEC} that the operator $\d$ allows to decompose
the exterior derivative $d$ as a BRS commutator.
This property, as already shown in~\cite{silvio}, gives an elegant and
simple procedure for solving the equations \equ{LADDER}.

Let us consider indeed the tower of descent equations which originates
from a local field polynomials $\O^{G}_{N}$ in the variables
$(e^{a}, \o^{a}_{~bm}, R^{a}_{~bmn}, T^{a}_{mn}, \th^{a}_{~b}, \h^{a})$
and their derivatives with ghost number $G$ and form degree $N$,
$N$ being the dimension of the space-time,
\bea
\label{TOWER}
&&s\O^{G}_{N}+d\O^{G+1}_{N-1}=0\non
&&s\O^{G+1}_{N-1}+d\O^{G+2}_{N-2}=0\non
&&~~~~~.....\non
&&~~~~~.....\non
&&s\O^{G+N-1}_{1}+d\O^{G+N}_{0}=0\non
&&s\O^{G+N}_{0}=0 \ ,
\eea
with $(\O^{G+1}_{N-1},...., \O^{G+N-1}_{1}, \O^{G+N}_{0})$ local
polynomials which, without loss of generality, will be always
considered as irreducible elements, i.e. they cannot be expressed as
the product of several factorized terms.
In particular, the ghost numbers $G=(0,1)$ correspond respectively to
an invariant gravitational Lagrangian and to an anomaly.

Thanks to the operator $\d$ and to the algebraic relations
\equ{DEC}-\equ{EXCOM}, in order to find a solution of the
ladder \equ{TOWER} it is sufficient to solve only the last equation
for the zero form $\O^{G+N}_{0}$.
It is easy to check that, once a nontrivial solution for $\O^{G+N}_{0}$
is known, the higher cocycles $\O^{G+N-q}_{q}$, $(q=1,...,N)$ are
obtained by repeated applications of the operator $\d$ on
$\O^{G+N}_{0}$, i.e.
\be
\label{HICO}
\O^{G+N-q}_{q}=\frac{\d^{q}}{q!}\O^{G+N}_{0}~~~,~~~q=1,...,N~~~,
{}~~~G=(0,1)~.
\ee
Let us emphasize also that solving the last equation of the tower
\equ{TOWER} is a problem of {\it local} BRS cohomology instead of a
modulo-$d$ one. One sees then that, by means of the decomposition
\equ{DEC}, the study of the cohomology of $s$ modulo $d$ is
reduced to the study of the local cohomology of $s$.
It is well-known indeed that, once a particular solution of the
descent equations \equ{TOWER} has been obtained, i.e. eq.\equ{HICO},
the search of the most general solution becomes essentially a problem
of local BRS cohomology.

Let us conclude this section by remarking that actually, also if a
fully characterization of the local cohomology of the BRS operator
in eqs.\equ{FORMTWO}-\equ{FORMZERO}
has not yet been obtained~\cite{next}, it is rather simple to produce some
interesting examples.
This is the aim of the next chapter.

\section{Some examples}

In this section we apply the previous algebraic setup,
eqs.\equ{TOWER}-\equ{HICO}, to discuss some explicit examples.
In particular we will focus on the cohomological origin of the
cosmological constant, of the Einstein and torsion Lagrangians
as well as of the Chern-Simons terms and of the gravitational
anomalies.
The analysis will be carried out for any space-time dimension, i.e.
the Lorentz group will be assumed to be $SO(N)$ with $N$ arbitrary.

\subsection{The cosmological constant}

The simplest local BRS invariant polynomial which one can define is
\be
\label{COSMOZERO}
\O^{N}_{0}=\frac{1}{N!}\ve_{a_{1}a_{2}.....a_{N}}
\h^{a_{1}}\h^{a_{2}}.....\h^{a_{N}}.
\ee
with $\ve_{a_{1}a_{2}.....a_{N}}$ the totally antisymmetric invariant
tensor of $SO(N)$.
Taking into account that in a $N$-dimensional space-time the product
of $(N+1)$ ghost fields $\h^{a}$ automatically vanishes, it is easily
checked that $\O^{N}_{0}$ identifies a cohomology class of
the BRS operator, i.e.
\be
s\O^{N}_{0}=0~~~,~~~\O^{N}_{0} \not= s\widehat{\O}^{N-1}_{0}.
\ee
The cocycle \equ{COSMOZERO} corresponds to the case $G=0$
(see eqs.\equ{TOWER}) and gives rise to the invariant
Lagrangian $\O^{0}_{N}$
\be
\label{COSMOCONST}
\O_{N}^{0}=\frac{\d^{N}}{N!}\O^{N}_{0}
=\frac{(-1)^{N}}{N!}\ve_{a_{1}a_{2}.....a_{N}}
e^{a_{1}}e^{a_{2}}.....e^{a_{N}},
\ee
which is easily recognized to coincide with the $SO(N)$ cosmological
constant. One sees thus that the cohomological origin of the
cosmological constant \equ{COSMOCONST} relies on the cocycles
\equ{COSMOZERO}.

\subsection{Einstein Lagrangians}

In this case, using the zero form curvature $R^{ab}_{~~mn}$, for the
cocycle $\O^{N}_{0}$ $(N>2)$ one gets
\be
\label{EINSTEINZERO}
\O^{N}_{0}=\frac{1}{2}\frac{1}{(N-2)!}\ve_{a_{1}a_{2}.....a_{N}}
R^{a_{1}a_{2}}_{~~~~mn}\h^{m}\h^{n}\h^{a_{3}}.....\h^{a_{N}},
\ee
to which it corresponds the term
\bea
\label{EINSTEINLAGR}
\O^{0}_{N}\=\frac{\d^{N}}{N!}\O^{N}_{0}\non
\=\frac{1}{2}\frac{(-1)^{N}}{(N-2)!}\ve_{a_{1}a_{2}.....a_{N}}
R^{a_{1}a_{2}}_{~~~~mn}e^{m}e^{n}e^{a_{3}}.....e^{a_{N}}\non
\=\frac{(-1)^{N}}{(N-2)!}\ve_{a_{1}a_{2}.....a_{N}}
R^{a_{1}a_{2}}e^{a_{3}}.....e^{a_{N}}.
\eea
Expression \equ{EINSTEINLAGR} is nothing but the Einstein Lagrangian
for the case of $SO(N)$.

Notice also that for the case of $SO(2)$ the
zero form cocycle $\O^{2}_{0}$
\be
\O^{2}_{0}=\frac{1}{2}\ve_{ab}R^{ab}_{~~mn}\h^{m}\h^{n}
\ee
turns out to be BRS-exact:
\be
\O^{2}_{0}=-s(\ve_{ab}\o^{ab}_{~~m}\h^{m}+\ve_{ab}\th^{ab}) \ .
\ee
As it is well-known, this implies that the two dimensional
Einstein Lagrangian
\be
\O^{0}_{2}=\ve_{ab}R^{ab}
\ee
is $d$-exact, i.e.
\be
\O^{0}_{2}=d(\ve_{ab}\o^{ab}).
\ee

\subsection{Generalized curvature Lagrangians}

Replacing in the Einstein Lagrangians \equ{EINSTEINLAGR} any pair of
vielbeins with the two form $R^{ab}$, we get another set of
gravitational Lagrangians containing higher powers of the
Riemann tensor.

To give an example, let us consider the zero form cocycle
\bea
\O^{2N}_{0}\=\frac{1}{(2N)!}\frac{1}{2^{N}}
(\ve_{a_{1}a_{2}a_{3}a_{4}.....a_{(2N-1)}a_{(2N)}}
R^{a_{1}a_{2}}_{~~~~b_{1}b_{2}}R^{a_{3}a_{4}}_{~~~~b_{3}b_{4}}
.....R^{a_{(2N-1)}a_{(2N)}}_{~~~~~~~~~~~~b_{(2N-1)}b_{(2N)}})\non
&\times&\!\!\!(\h^{b_{1}}\h^{b_{2}}\h^{b_{3}}\h^{b_{4}}.....
\h^{b_{(2N-1)}}\h^{b_{(2N)}}) \ .
\eea
Using eq.\equ{HICO}, for the corresponding invariant Lagrangian
one gets
\bea
\O^{0}_{2N}\=\frac{\d^{(2N)}}{(2N)!}\O^{2N}_{0}\non
\=\frac{1}{(2N)!}\frac{1}{2^{N}}
(\ve_{a_{1}a_{2}a_{3}a_{4}.....a_{(2N-1)}a_{(2N)}}
R^{a_{1}a_{2}}_{~~~~b_{1}b_{2}}R^{a_{3}a_{4}}_{~~~~b_{3}b_{4}}
.....R^{a_{(2N-1)}a_{(2N)}}_{~~~~~~~~~~~~b_{(2N-1)}b_{(2N)}})\non
&\times&\!\!\!(e^{b_{1}}e^{b_{2}}e^{b_{3}}e^{b_{4}}.....
e^{b_{(2N-1)}}e^{b_{(2N)}})\non
\=\frac{1}{(2N)!}
\ve_{a_{1}a_{2}a_{3}a_{4}.....a_{(2N-1)}a_{(2N)}}
R^{a_{1}a_{2}}R^{a_{3}a_{4}}.....R^{a_{(2N-1)}a_{(2N)}} \ .
\eea

\subsection{Lagrangians with torsion}

It is known that~\cite{baulieu},
for special values of the space-time dimension
$N$, i.e. $N=(4M-1)$ with $M\ge1$, there is the possibility of defining
nontrivial invariant Lagrangians which explicitly contain the
torsion.
\newline
Let us begin by considering first the simpler case of $SO(3)$ $(M=1)$.
By making use of the zero form $T^{a}_{mn}$, for the cocycle
$\O^{3}_{0}$ one has\footnote{Tangent space indices are rised and
lowered with the flat metric $g_{ab}$, $\h_{a}=g_{ab}\h^{b}$.}
\be
\O^{3}_{0}=\frac{1}{2}T^{a}_{mn}\h^{m}\h^{n}\h_{a} \ ,
\ee
from which one gets the three dimensional torsion Lagrangian
\bea
\O^{0}_{3}=\frac{\d^{3}}{3!}\O^{3}_{0}
\=-\frac{1}{2}T^{a}_{mn}e^{m}e^{n}e_{a}
=-T^{a}e_{a} \ .
\eea
Generalizing to the case of $SO(4M-1)$ with $(M>1)$, one finds
\bea
\O^{4M-1}_{0}\=\frac{1}{2^{(2M-1)}}(T_{k~\!\!m_{1}m_{2}}
R^{k}_{~a_{1}m_{3}m_{4}}R^{a_{1}}_{~~a_{2}m_{5}m_{6}}....
R^{a_{(2M-3)}}_{~~~~~~~a_{(2M-2)}m_{(4M-3)}m_{(4M-2)}})\non
&\times&\!\!\!(\h^{m_{1}}\h^{m_{2}}....\h^{m_{(4M-3)}}\h^{m_{(4M-2)}}
\h^{a_{(2M-2)}}) \ ,
\eea
which yields the following torsion Lagrangians
\bea
\O^{0}_{4M-1}\=-\frac{1}{2^{(2M-1)}}(T_{k~\!\!m_{1}m_{2}}
R^{k}_{~a_{1}m_{3}m_{4}}R^{a_{1}}_{~~a_{2}m_{5}m_{6}}....
R^{a_{(2M-3)}}_{~~~~~~~a_{(2M-2)}m_{(4M-3)}m_{(4M-2)}})\non
&\times&\!\!\!(e^{m_{1}}e^{m_{2}}....e^{m_{(4M-3)}}e^{m_{(4M-2)}}
e^{a_{(2M-2)}})\non
\=-T_{k}R^{k}_{~a_{1}}R^{a_{1}}_{~~a_{2}}....
R^{a_{(2M-3)}}_{~~~~~~~a_{(2M-2)}}
e^{a_{(2M-2)}} \ .
\eea

Let us mention also the possibility of defining invariant torsion terms
which are polynomial in $T^{a}_{mn}$. These Lagrangians exist in any
space-time dimension and are easily obtained from the
$SO(N)$ zero form cocycle
\be
\label{POLTORS}
\O^{N}_{0}=\frac{1}{N!}\ve_{a_{1}a_{2}.....a_{N}}
\h^{a_{1}}\h^{a_{2}}.....\h^{a_{N}} \PP(T) \ ,
\ee
with $\PP(T)$ a scalar polynomial in the torsion as,
for instance~\cite{kumm} (see also~\cite{sez} for generalization),
\be
  \PP(T) = T^{a}_{mn} T_{a}^{mn} \ .
\label{PTEXAMPLE}
\ee
The corresponding invariant torsion Lagrangians are given then by
\be
\label{PTLAGRANG}
\O_{N}^{0}=\frac{\d^{N}}{N!}\O^{N}_{0}
=\frac{(-1)^{N}}{N!}\ve_{a_{1}a_{2}.....a_{N}}
e^{a_{1}}e^{a_{2}}.....e^{a_{N}}\PP(T) \ .
\ee

\subsection{Chern-Simons terms and anomalies}

For what concerns the Chern-Simons terms and the Lorentz and
diffeomorphism anomalies (see also Appendix B for the so-called
first family diffeomorphism cocycles~\cite{tonin1,tonin2,tonin3})
we recall that, as mentioned
in the introduction, an algebraic analysis based on the decomposition
\equ{DEC} has been recently carried out by~\cite{werneck}.

Let us remark, however, that the decomposition found in~\cite{werneck}
gives
rise to a commutation relation between the operators $\d$ and $d$ which
contrary to the present case (see eq.\equ{EXCOM}) does not vanish.
This implies the existence of a further operator $\GG$ of degree one
which has to be taken into account in order to solve the
ladder \equ{TOWER}.

Actually, the existence of the operator $\GG$ relies on the fact that
the decomposition of the exterior differential $d$ found
in~\cite{werneck}
does not take into account the explicit presence of the vielbein
$e^{a}$ and of the torsion $T^{a}$. It holds for a functional space
whose basic elements are built only with the spin connection
$\o^{a}_{~b}$ and the Riemann tensor $R^{a}_{~b}$, this choice
being sufficient to characterize all known Lorentz anomalies and
related second family diffeomorphism cocycles~\cite{tonin1,tonin2}.

It is remarkable then to observe that the algebra between $s$, $\d$,
and $d$ gets simpler only when the vielbein $e^{a}$ and the torsion
$T^{a}$ are naturally present.
Let us emphasize indeed that the particular elementary form of the
operator $\d$ in eq.\equ{DECETA} is due to the use of the
tangent space ghost $\h^{a}$ whose introduction requires explicitly
the presence of the vielbein $e^{a}$.

For the sake of clarity and to make contact with the results obtained
in~\cite{werneck}, let us discuss in details the construction of the
$SO(3)$ Chern-Simons term. In this case the tower \equ{TOWER} takes
the form
\bea
&&s\O^{0}_{3}+d\O^{1}_{2}=0\non
&&s\O^{1}_{2}+d\O^{2}_{1}=0\non
&&s\O^{2}_{1}+d\O^{3}_{0}=0\non
&&s\O^{3}_{0}=0 \ ,
\eea
where, according to eq.\equ{HICO},
\bea
&&\O^{2}_{1}=\d\O^{3}_{0}\non
&&\O^{1}_{2}=\frac{\d^{2}}{2!}\O^{3}_{0}\non
&&\O^{0}_{3}=\frac{\d^{3}}{3!}\O^{3}_{0} \ .
\eea
In order to find a solution for $\O^{3}_{0}$ we use the redefined
Lorentz ghost
\be
\widehat{\th}^{a}_{~b}=\o^{a}_{~bm}\h^{m}+\th^{a}_{~b} \ ,
\ee
which, from eq.\equ{DECETA}, transforms as
\be
\d\widehat{\th}^{a}_{~b}=-\o^{a}_{~b} \ .
\ee
For the cocycle $\O^{3}_{0}$ one gets then
\be
\O^{3}_{0}=\frac{1}{3!}\widehat{\th}^{a}_{~b}
\widehat{\th}^{\hspace{0.04cm}b}_{~c}
\widehat{\th}^{c}_{~a}-\frac{1}{4}R^{a}_{~bmn}\h^{m}\h^{n}
\widehat{\th}^{\hspace{0.04cm}b}_{~a} \ ,
\ee
from which $\O^{2}_{1}$, $\O^{1}_{2}$, and $\O^{0}_{3}$ are
computed to be
\bea
\O^{2}_{1}=-\frac{1}{2}\o^{a}_{~b}
\widehat{\th}^{\hspace{0.04cm}b}_{~c}
\widehat{\th}^{c}_{~a}
+\frac{1}{2}R^{a}_{~bmn}e^{m}\h^{n}
\widehat{\th}^{\hspace{0.04cm}b}_{~a}
+\frac{1}{4}R^{a}_{~bmn}\h^{m}\h^{n}\o^{b}_{~a} \ ,
\eea
\bea
\label{CHERNTWO}
\O^{1}_{2}=\frac{1}{2}(\o^{a}_{~b}\o^{b}_{~c}\widehat{\th}^{c}_{~a}
-R^{a}_{~b}\widehat{\th}^{\hspace{0.04cm}b}_{~a}
-R^{a}_{~bmn}e^{m}\h^{n}\o^{b}_{~a}) \ ,
\eea
\bea
\label{CHERNTHREE}
\O^{0}_{3}=\frac{1}{2}(R^{a}_{~b}\o^{b}_{~a}
-\frac{1}{3}\o^{a}_{~b}\o^{b}_{~c}\o^{c}_{~a}) \ .
\eea
In particular, expression \equ{CHERNTHREE} gives the familiar $SO(3)$
Chern-Simons gravitational term. Finally, let us remark that the
cocycle $\O^{1}_{2}$ of eq.\equ{CHERNTWO}, when referred to $SO(2)$,
reduces to the expression
\be
-\frac{1}{2}(d\o^{a}_{~b})\th^{b}_{~a}
\ee
which directly gives the two dimensional Lorentz anomaly.

\section{The geometrical meaning of the operator $\d$}

Having discussed the role of the operator $\d$ in finding explicit
solutions of the descent equations \equ{TOWER}, let us turn now to
the study of its geometrical meaning.
As we shall see, this operator turns out to possess a quite simple
geometrical interpretation which will reveal an unexpected and so far
unnoticed elementary structure of the ladder \equ{TOWER}.

Let us begin by observing that all the cocycles $\O^{G+N-p}_{p}$
$(p=0,...,N)$ entering the descent equations \equ{TOWER} are of the
same degree (i.e. $(G+N)$), the latter being given by the sum of
the ghost number and of the form degree.

We can collect then, following~\cite{tataru}, all the $\O^{G+N-p}_{p}$
into a unique cocycle $\widehat{\O}$ of degree $(G+N)$ defined as
\be
\widehat{\O}=\sum_{p=0}^{N}\O^{G+N-p}_{p} \ .
\ee
This expression, using eq.\equ{HICO}, becomes
\be
\label{SUM}
\widehat{\O}=\sum_{p=0}^{N}\frac{\d^{p}}{p!}\O^{G+N}_{0} \ ,
\ee
where the cocycle $\O^{G+N}_{0}$, according to its form degree,
depends only on the set of zero form variables
$(\o^{a}_{~bm}, R^{a}_{~bmn}, T^{a}_{mn}, \th^{a}_{~b}, \h^{a})$
and their tangent space derivatives $\6_{m}$.
Taking into account that under the action of the operator $\d$ the
form degree and the ghost number are respectively rised and lowered
by one unit and that in a space-time of dimension $N$ a $(N+1)$-form
identically vanishes, it follows that eq.\equ{SUM} can be rewritten
in a more suggestive way as
\be
\label{EXPONENT}
\widehat{\O}=e^{\d}\O^{G+N}_{0}(\h^{a}, \th^{a}_{~b}, \o^{a}_{~bm},
R^{a}_{~bmn}, T^{a}_{mn}) \ .
\ee
Let us make now the following elementary but important remark.
As one can see from eq.\equ{DECETA}, the operator $\d$ acts as a
translation on the ghost $\h^{a}$ with an amount given by $(-e^{a})$.
Therefor $e^{\d}$ has the simple effect of shifting $\h^{a}$ into
$(\h^{a}-e^{a})$. This implies that the cocycle \equ{EXPONENT} takes
the form
\be
\label{SHIFT}
\widehat{\O}=\O^{G+N}_{0}(\h^{a}-e^{a}, \th^{a}_{~b}, \o^{a}_{~bm},
R^{a}_{~bmn}, T^{a}_{mn}) \ .
\ee
This formula collects in a very elegant and simple expression the
solution of the descent equations \equ{TOWER}.

In particular, it states the important result that:

\begin{quote}
 {\it To find a nontrivial solution of the ladder} \equ{TOWER}
{\it it
is sufficient to replace the variable $\h^{a}$ with $(\h^{a}-e^{a})$
in the zero form cocycle $\O^{G+N}_{0}$ which belongs to the local
cohomology of the BRS operator $s$. The expansion
of $\O^{G+N}_{0}(\h^{a}-e^{a}, \th^{a}_{~b}, \o^{a}_{~bm},
R^{a}_{~bmn}, T^{a}_{mn})$ in powers of the one form vielbein
$e^{a}$ yields then all the searched cocycles $\O^{G+N-p}_{p}$.}
\end{quote}

It is a simple exercise to check now that all the invariant
Lagrangians and Chern-Simons terms computed in the previous section
are indeed recovered by simply expanding the corresponding zero form
cocycles $\O^{G+N}_{0}$ taken as functions of $(\h^{a}-e^{a})$.

Let us conclude by remarking that, up to our knowledge, expression
\equ{SHIFT} represents a deeper understanding of the algebraic
properties of the gravitational ladder \equ{TOWER} and of the role
played by the vielbein $e^{a}$ and the associated ghost $\h^{a}$.

\section*{Conclusion}

The algebraic structure of gravity with torsion has been analyzed
in the context of the Maurer-Cartan horizontality formalism by
introducing an operator $\delta$ which allows
to decompose the exterior space-time derivative as a BRS commutator.
Such a decomposition gives a simple and elegant way of solving
the Wess-Zumino consistency condition corresponding to invariant
Lagrangians and anomalies. The same technique can be
applied to the study of the gravitational coupling of Yang-Mills
gauge theories as well as to the characterization of
the Weyl anomalies~\cite{next}.

\section*{Appendices}

Appendix A is devoted to the computation of some commutators
involving the tangent space derivative $\6_{a}$ introduced in
Section 2.

In the Appendix B we show how the so-called first family
diffeomorphism anomalies can be recovered with the help of
the decomposition \equ{DEC}.

\subsection*{A~~~Commutation relations}

\setcounter{equation}{0}
\renewcommand{\theequation}{A.\arabic{equation}}

In order to find the commutator involving two tangent space derivatives
$\6_{a}$, we make use of the fact that the usual space-time derivatives
$\6_{\mu}$ have vanishing commutator:
\be
[\6_{\mu},\6_{\nu}]=0 \ .
\ee
{}From
\be
\6_{\mu}=e^{m}_{\mu}\6_{m}
\ee
one gets
\bea
[\6_{\mu},\6_{\nu}]=0 \= [e^{m}_{\mu}\6_{m},e^{n}_{\nu}\6_{n}]\non
\= e^{m}_{\mu}e^{n}_{\nu}[\6_{m},\6_{n}]
+e^{m}_{\mu}(\6_{m}e^{n}_{\nu})\6_{n}
-e^{n}_{\nu}(\6_{n}e^{m}_{\mu})\6_{m}\non
\=e^{m}_{\mu}e^{n}_{\nu}[\6_{m},\6_{n}]
+(\6_{\mu}e^{k}_{\nu}-\6_{\nu}e^{k}_{\mu})\6_{k}\non
\=e^{m}_{\mu}e^{n}_{\nu}[\6_{m},\6_{n}]
+(T^{k}_{\mu\nu}-\o^{k}_{~n\mu}e^{n}_{\nu}
+\o^{k}_{~m\nu}e^{m}_{\mu})\6_{k}\non
\=e^{m}_{\mu}e^{n}_{\nu}\{[\6_{m},\6_{n}]
+(T^{k}_{mn}-\o^{k}_{~nm}+\o^{k}_{~mn})\6_{k}\} \ ,
\eea
so that
\be
[\6_{m},\6_{n}]=-(T^{k}_{mn}+\o^{k}_{~mn}-\o^{k}_{~nm})\6_{k} \ .
\ee
\newline
For the commutator of $d$ and $\6_{m}$ we get
\bea
[d,\6_{m}]\=[e^{n}\6_{n},\6_{m}]\non
\=-(\6_{m}e^{k})\6_{k}-e^{n}[\6_{m},\6_{n}]\non
\=-(\6_{m}e^{k})\6_{k}+e^{n}(T^{k}_{mn}+\o^{k}_{~mn}
-\o^{k}_{~nm})\6_{k} \ .
\eea
Analogously, from
\be
[s,\6_{\mu}]=0
\ee
one easily finds
\bea
[s,\6_{m}]\=(\6_{m}\h^{k}-\th^{k}_{~m})\6_{k}
+\h^{n}[\6_{m},\6_{n}]\non
\=(\6_{m}\h^{k}-\th^{k}_{~m}-T^{k}_{mn}\h^{n}-\o^{k}_{~mn}\h^{n}
+\o^{k}_{~nm}\h^{n})\6_{k} \ .
\eea

\subsection*{B~~~First family diffeomorphism anomalies}

\setcounter{equation}{0}
\renewcommand{\theequation}{B.\arabic{equation}}

In this Appendix we give a brief discussion on the first family
diffeomorphism anomalies.

As it is well-known, the diffeomorphism cocycles can be splitted
in two groups, usually referred as first and second family
anomalies. The latters are defined by the following result,
valid for any space-time dimension:

{\bf Diffeomorphism anomalies}~\cite{tonin1,tonin2}

\begin{quote}
On the space of local polynomials in the connection $\omega$,
the vielbein $e$ and its inverse $E$, the Riemann tensor $R$
and the torsion $T$, the most general diffeomorphism
anomaly $\AA_{diff}$ has the form
\be
\label{FAMILY}
\AA_{diff}=\int d^{N}\!x(b^{\mu\nu}_{\s}\6_{\mu}\6_{\nu}\x^{\s}
+b\6_{\mu}\x^{\mu}),
\ee
where $b^{\mu\nu}_{\s}$ is a tensor under linear
$GL(N)$-transformations
and $b$ is a scalar density which cannot be written as a total
derivative, i.e.
\be
\label{SD}
b=e\MM~~~,~~~b\not=\6_{\mu}\hat{b}^{\mu},
\ee
with $\MM$ a scalar quantity
\be
\label{MM}
s\MM=-\x^{\l}\6_{\l}\MM \ .
\ee
\end{quote}

The variables $\x^{\mu}$ and $e$ in eqs.\equ{FAMILY}, \equ{SD} denote
respectively the diffeomorphism ghost of eq.\equ{ETA}
and the determinant of the vielbein $e^{a}_{\mu}$
\bea
e\=det~e^{a}_{\mu}=\frac{1}{N!}\ve_{a_{1}a_{2}.....a_{N}}
\ve^{\mu_{1}\mu_{2}.....\mu_{N}}e^{a_{1}}_{\mu_{1}}e^{a_{2}}_{\mu_{2}}
.....e^{a_{N}}_{\mu_{N}} \ ,\non
se\=-\6_{\l}(\x^{\l}e) \ .
\eea
Let us remark also that
\be
\x^{\l}\6_{\l}=\h^{m}\6_{m} \ ,
\ee
so that eq.\equ{MM} becomes
\be
s\MM=-\h^{m}\6_{m}\MM \ .
\ee

The coefficients $b$ and $b^{\mu\nu}_{\s}$ in eq.\equ{FAMILY} define
the so-called first and second family diffeomorphism anomalies and
are related, through a nonpolynomial Wess-Zumino action~\cite{zumino},
respectively to Weyl and Lorentz anomalies.

Since the second family diffeomorphism anomalies have been already
extensively discussed in~\cite{werneck}, let us focus here on the
first family cocycles.
Actually, it is not difficult to verify that the cohomological origin
of the first family diffeomorphism anomalies relies on the zero form
\be
\O^{N+1}_{0}=\frac{1}{N!}\ve_{a_{1}a_{2}.....a_{N}}
\h^{a_{1}}\h^{a_{2}}.....\h^{a_{N}}(\6_{m}\h^{m})\MM \ ,
\ee
which is easily seen to be BRS invariant.

Indeed, according to the general procedure, i.e. equation \equ{HICO},
the corresponding anomaly is given by
\bea
\O^{1}_{N}\=\frac{1}{N!}\d^{N}\O^{N+1}_{0}\non
\=\frac{(-1)^{N}}{N!}\ve_{a_{1}a_{2}.....a_{N}}
e^{a_{1}}e^{a_{2}}.....e^{a_{N}}(\6_{m}\h^{m})\MM\non
\+\frac{(-1)^{N}}{(N-1)!}\ve_{a_{1}a_{2}.....a_{N}}
\h^{a_{1}}e^{a_{2}}.....e^{a_{N}}(\6_{m}e^{m})\MM \ .
\eea
Expressing the tangent space ghost $\h^{a}$ in terms of the
diffeomorphism ghost $\x^{\mu}$ (see eq.\equ{ETA}) and using the
identity
\bea
e^{1}.....e^{N}\=\frac{1}{N!}\ve_{a_{1}.....a_{N}}
e^{a_{1}}.....e^{a_{N}}\non
\=\frac{1}{N!}\ve_{a_{1}.....a_{N}}e^{a_{1}}_{\mu_{1}}.....
e^{a_{N}}_{\mu_{N}}
dx^{\mu_{1}}.....dx^{\mu_{N}}\non
\=\frac{1}{N!}\ve_{a_{1}.....a_{N}}\ve^{\mu_{1}.....\mu_{N}}
e^{a_{1}}_{\mu_{1}}.....e^{a_{N}}_{\mu_{N}}
dx^{1}.....dx^{N}\non
\=ed^{N}\!x \ ,
\eea
for $\O^{1}_{N}$ one gets
\be
\label{FIRSTFAMILYLAGR}
\O^{1}_{N}=e\MM(\6_{\l}\x^{\l})d^{N}\!x \ .
\ee
We have thus recovered the first family anomalies by means of the
operator $\d$.
Finally, as already mentioned, let us recall that the cocycle
\equ{FIRSTFAMILYLAGR}, also if cohomologically nontrivial in the space
of polynomials, can be mapped into a Weyl
anomaly~\cite{tonin2,tonin3,bast,deser} by means of
a nonpolynomial action. Indeed, using the logarithm of the
determinant of the vielbein as the Goldstone boson field~\cite{zumino},
one easily checks that
\be
\int d^{N}\!x~e\MM(\6_{\l}\x^{\l})=-s\int d^{N}\!x~e\MM(log~e) \ .
\ee


\end{document}